# REFLECTIONS ON THE NATURE OF GENIUS: ON THE 300TH ANNIVERSARY OF MIKHAIL LOMONOSOV (1711-1765)*

V.Shiltsev[#], FNAL, Batavia, IL 60510, U.S.A.

*Abstract*

This presentation goes beyond celebratory narration of the life and scientific achievements of Russia's first modern scientist Mikhail Vasilievich Lomonosov (1711-1765) [1,2,3]. Coming from the notion of complexity of sciences [4], we introduce "a genius formula" *G=TBD* for semi-qualitative evaluation of a person's impact on the society, distinguish two type of geniuses, give several examples and draw general conclusions. The work largely follows presentation at the Fermilab Colloquium in November of 2011 [5].

## MIKHAIL LOMONOSOV

Mikhail Lomonosov was born November 19, 1711 into the family of a relatively free "state peasant"-turned-fisherman in a Northern Russian village near Archangel. In pursuit of opportunity he escaped from home at the age of 19 to reach Moscow by feet over 800 miles of winter roads and enter Zaikono-Spassky academy.

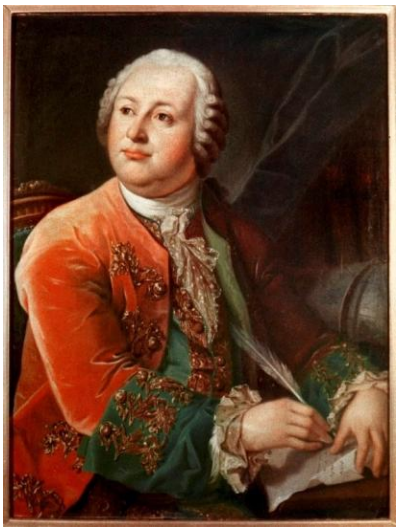

Figure 1: Portrait of Mikhail Lomonosov (1711-1765)

Half-starving on a stipend of 3 kopeks a day, in just 4 years he finished an 8 year course in Latin, Greek, Church Slavonic, geography, history, philosophy and the Catechism. From there he was sent to Sankt Petersburg Academy of Sciences ("the Academy") to continue his education among the 12 best students in 1736. The same year he was sent by the Academy to University of Marburg to Germany to further study mathematics, chemistry, mining, natural history, physics, mechanics, hydraulics, and humanities with Christian Wolff (1679-1754)-a renowned encyclopedic scientist and philosopher, and a key follower of Leibniz–who came to highly regard Lomonosov's abilities. He returned to Russia in 1741 where on the merits of his numerous excellent scientific reports regularly sent from abroad and a glorious poetic ode to Empress Anna, he received an appointment as an Adjunct of Physics in the St. Petersburg Academy (at that time totally dominated by foreign scholars). He was the first native-born Russian Academician elected in 1745 and served as a member of Academy's Chancellery, in charge of all scientific and educational activities and departments, from 1757 till his death on April 15, 1765. Over years Lomonosov fiercely fought the Academy's decline, trying to get it back on the track set by Peter the Great. He succeeded in this challenge by increasing the number of scientific publications in Russian (in addition to Latin and German), and by insisting the Academicians deliver regular lectures in Russian. The result was a significantly increased number of Russian academicians as well as interns and students in the Academy's Gymnasium. In 1755 he founded Russia's first University in Moscow, now named after him. Lomonosov was elected an honorary member of the Swedish Academy of Sciences (1760), the St. Petersburg Academy of Arts (1763), and a member of the Bologna Academy of Sciences(1764).

The polymathic nature of this titan of the Russian Enlightenment can be gleaned from the content of his *Complete Works* [6]: vols. 1-4–works on physics, chemistry, astronomy; vol. 5–mineralogy, metallurgy and geology, vol. 6–Russian history, economics and geography, vols. 7-8–philology, poetry, prose, vols. 9-11–correspondence, letters and translations. The depth of his insights is even more remarkable. Just in natural sciences alone, Lomonosov performed by himself more than 4000 chemical tests in Russia's first national laboratory and championed explanations of all physical and chemical phenomena on the basis of corpuscular mechanics in a continuous ether; he coined the term "physical chemistry" in 1752 and thought of absolute cold as a condition where the corpuscles ceased their linear and rotational motions.

17 years prior to analogous results by A.Lavoisier, Lomonosov experimentally proved the law of conservation of matter by showing that lead plates in a sealed vessel without access to air do not change their weight after heating (1756); based on the results of the first quantitative experimental studies of electricity in 1744-1756 –which were quite dangerous as his colleague Georg Richmann was killed by ball lightning and

_________________
*invited talk at the CERN's "Chamonix-2012" workshop; work in part supported by FRA LLC which operates Fermilab under contract No. DE-AC02-07CH11359 with the U.S. Department of Energy
[#]shiltsev@fnal.gov

Lomonosov himself "miraculously survived"– he proposed an original theory of atmospheric electricity that went beyond B.Franklin's, and explained with it lightning and the polar lights. Looking for a way to send meteorological instruments and electrometers aloft, he designed and built the first working helicopter model (1754). This used two propellers rotating in opposite directions for torque compensation and, powered by a clock spring, managed to provide some 10 grams of lift.

During the transit of Venus on May 26, 1761 Lomonosov discovered the atmosphere of Venus by observing a bright aureole around the planet at the ingress and egress, and gave a detailed optical explanation of the effect by refraction. Thirty years before Herschel, in 1762, he invented and built a practical reflector telescope of a new type with the primary mirror tilted by 4 degrees so one could view the formed image directly in a side eyepiece; later that same year he invented a siderostat mechanism which allowed tracking of the stars by tilting a flat mirror in front rather than the entire telescope. See recent publications [1,2,3] for more detail and examples.

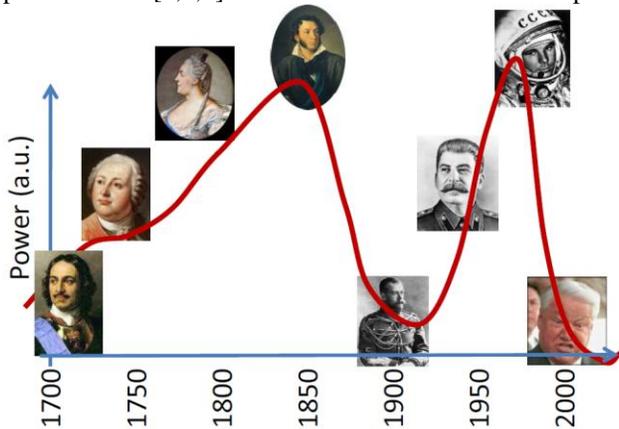

Figure 2: Lomonosov's place in modern Russian history.

Lomonosov is widely recognized as the foremost name in Russian culture and history – see Fig.2 which cartoonishly illustrates the passionate power of Russian ethos over the past three centuries. Pushkin concluded on him "…*between Peter I and Catherine II, he (Lomonosov) was the original champion of the Enlightenment. He founded the first University: better to say, he himself was our first University…*" [7]. His tercentennial is being celebrated statewide in Russia in 2011. Lomonosov was, however, not well known in the Western scientific circles because among his contemporaries, the enormous breadth of his achievements, e.g., his works in grammar, mosaic art and especially poetry, outshone his work in Natural Philosophy. The lack of awareness was also due to the weak national scientific community till the late 1800's, to the lack of personal contacts with the West (except Euler), and, partly, to his relatively short life.

## GENIUS FORMULA

It was recently shown that many complex scientific and technical systems demonstrate exponential performance progress with time :

$$Performance \propto \exp(T/C) \quad (1)$$

(see, e.g. [4]). The characteristic progress time $C$ - a measure of empirical difficulty and complexity – is analyzed for particle colliders, astrophysical searches for galaxies and exoplanets, protein structure determination and compared with computers and thermonuclear fusion reactors – see Figs.3 and 4 below.

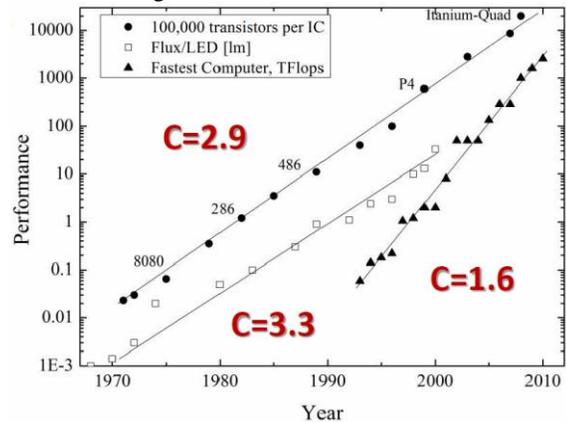

Figure 3: Progress of the fastest computers (triangles), light output per LED (squares) and number of transistors per chip ("Moores' law", circles) (from [4]).

The underlying explanation for the exponential performance progress is the fact that the complex systems have many intervened parameters and problems which can not be solved all at once, hence, addressed step by step. The the goal and outcome for each improvements step is certain percentage ($m$-percent) increase or $x$-fold increase with respect to what is already achieved, so after $n$ steps, the performance is either $(1+m/100)^n \approx e^{nm/100}$ or $x^n = e^{n \ln(x)}$. Regular implementation of the fractional steps leads to the exponential growth.

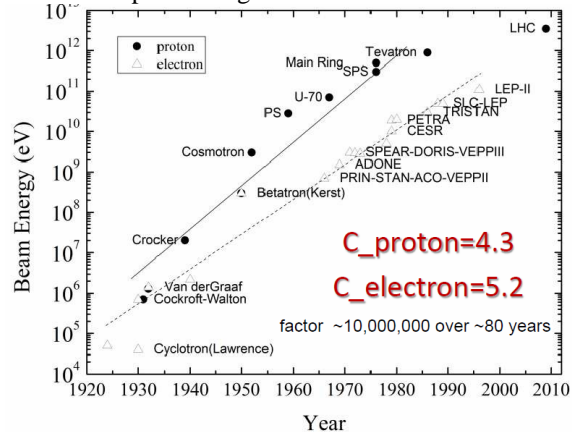

Figure 4: Another illustration of the exponential performance progress in complex systems – maximum achieved beam energy in particle accelerators ("Livingston plot", from [4]).

Coefficients *C* (complexities) of several systems are calculated and given below [4]:

|  | C (years) | Interval |
|---|---|---|
| Supercomputers | **1.6 ±0.1** | 1993-2010 |
| Fusion reactors | **2.4 ±0.2** | 1969-1999 |
| Transistors/IC | **2.7 ±0.05** | 1971-2009 |
| Galaxies surv'd | **3.0 ±0.1** | 1985-1990 |
| Light per LED | **3.3 ±0.1** | 1969-2000 |
| Laser power | **3.3 ±0.5** | 1975-2000 |
| Protein structures | **4.2 ±0.2** | 1976-2010 |
| Exoplanets search | **4.2 ±0.3** | 1991-2010 |
| Energy accelerators | **5.2 ±0.3** | 1930-1990 |

A natural question is "why and how do these extended periods of the fast (exponential) growth begin?" We introduce the concept of "genius" to answer that question, namely, the "genius effect" which drastically changes the pace of a system's evolution over relatively short period of time – usually, associated with discovery, invention or a deep insight (see Fig.5).

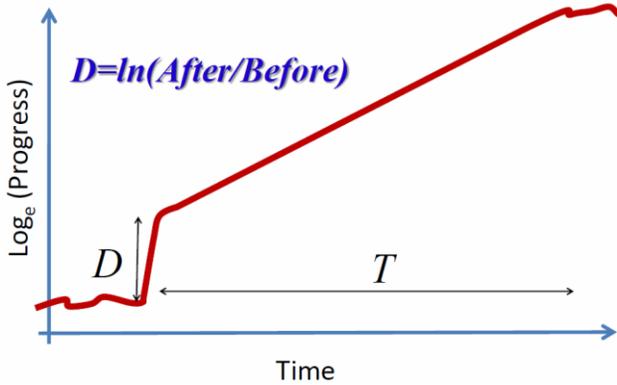

Figure 5: Schematic representation of the "genius effect" which leads to step-like change in the progress due to profound innovation or discovery.

The scale of the initial impact is given by $D=\ln(After/Before)$. If the impact of the scientist lasted over period of time *T*, one can introduce the *genius coefficient G*:

$$G = \ln(T) \times B \times D \qquad (2)$$

where additional factor *B* takes into account breadth of the scientist (sum up over the number of areas where he/she contributed to). Scientists with $B>1$ are rare and called "polymaths", famous examples are Leonardo, Newton and Poincare. Mikhail Lomonosov was such a polymath with $G=7 \pm 1.5$ – see below the table of his contributions to varios field of science and art (rough estimates):

|  | $ln(T) \times D$ | G |
|---|---|---|
| Chemistry | $ln(20) \times 0.1$ | 0.3 ±0.1 |
| Russian poetry | $ln(60) \times 1/3$ | 1.3 ±0.5 |
| Venus atmosphere | $ln(60) \times 1/5$ | 0.8 ±0.3 |
| Optics | $ln(100) \times 1/20$ | 0.2 ±0.1 |
| Russian history | $ln(250) \times 1/10$ | 0.5 ±0.2 |
| Geography | $ln(60) \times 1/4$ | 0.9 ±0.3 |
| Russian grammar | $ln(100yrs) \times ln(2)$ | 3 ±1 |
| **TOTAL** |  | **7.0 ±1.5** |

For comparison, one can estimate genius coefficients of:
- Good, solid scientists        $G=0.01-0.03$
- Great inventors               $G=0.1-2$
- Nobel laureates               $G=0.3-4$
- Aristotle, Galileo, Newton, Lomonosov  $G=4-9$
- Shakespeare, Pushkin          $G=7-12$

It's interesting to note that, some 200 modern day sciences altogether generate some $G \approx 4-10$ units a year - such estimate of the *Global Genius Product* (GGP) comes from average complexity coefficient $C \approx 20-50$ -see Eq.(1). In principle, the "genius coefficient" formula Eq.(2) can be applied to other entities (beyond humans) – e.g. the God of Bible, who created the world in 6 days some 7519 years ago would have $G \approx 120$, while the Big Bang act of the Universe start up 14 billion year ago is equivalent to $G \approx 300-600$.

## DISCUSSION

The "genius formula" proposed above Eq.(2) should be used with caution, as for comparative studies the base of estimates *Before* and *After* must be properly chosen (e.g. it is easy to get high values of *G* in local reference systems). Also, it is not applicable in the not-well-quantifiable systems like art, beauty or music.

On the other hand, Eq.(2) reflects many features a scientist would agree with. For example, our perception of times is undoubtedly logarithmic: for us "now" means less than 1-3 years old, "recently" is for things from 3-10 years ago, "modern" attributes to events 10-30 years back, "golden age" points to 30-100 year ago, "classic science" refers to 100-300 years old, and "ancient" is all that older than 300-1000 years.

Logarithmic scale of the impact $D=\ln(After/Before)$ is not novelty at all. Half a century ago Russian Nobel Prize laureate Lev Landau classified theoretical physicists according to their achievement using a logarithmic scale [8]. According to his ranking system, a member of the lower class achieved 10 times less than a member of the preceding class. He placed Einstein in ½ class. In the 1st class he placed Bohr, Dirac, Heisenberg, Schrödinger, and Fermi. Thus, he thought that Einstein contributed to physics $10^{1/2} \approx 3$ times more than Dirac or Schrödinger. Landau originally ranked himself as a 2.5 but later promoted to a 2 [9].

One could also add, that large impact imposed by great geniuses on the society has been always accompanied by remarkable stories – sometimes not even true - which deeply resonate in us and leave long lasting memories. Examples are many, the most widespread are Archimedes *Eureka moment,* Galileo's *Eppur si muove (Still it moves!),* Newton's *apple*, Ben Franklin's *kite experiment*, Lomonosov's trail to Moscow with frozen fish sleigh convoy, Mendeleev's discovery of the *Periodic Table* while sleeping. Among modern ones – Richard Feynman's *O-ring moment*, the story of Stephen Hawking and the story of Russian mathematician Grigori Perelman who proved *Poincare's Conjecture* in 2002 and since then declined all the honors, prizes and awards. "Creative disobedience of genius" often adds to the fame as noted by Petr Kapitsa [10].

## CONCLUSIONS

We can clearly see that the system of the humankind is very complex - both in terms of hierarchy of connections and *C*-coefficients Eq.(1). Obviously, it is progressing and moving somewhere as it generates new knowledge, ideas, arts, inventions, etc. It's not fully clear yet how the world's *GGP* - global genius product in the sense of the "genius coefficient" Equation (2) – is being created, but reflecting back it is well realized how hard is to get to new heights and how easy we could lose positions. To avoid the loss, an advanced (scientific) society is needed as it leads to appearance of genius(es). We also know examples of enormous transformative power geniuses exert on the society.

To understand who we are and where are going, we must comprehend our geniuses of a type of Galileo, Newton, Einstein, Leibniz, Franklin, and Lomonosov. With that, I congratulate all of us with 300[th] anniversary of Mikhail Lomonosov!